\documentclass[showpacs,preprintnumbers,amsmath,twocolumn,amssymb,prb]{revtex4}
\usepackage{verbatim} 
\usepackage{dcolumn}
\usepackage{float}
\usepackage{caption}
\usepackage{bm}
\usepackage{graphicx}
\usepackage{amssymb}
\usepackage{subfig}
\usepackage{color}
\begin{document}

\title{Dephasing of a Qubit due to Quantum and Classical Noise}
\author{Ebad Kamil}
\email{kamil@theorie.physik.uni-goettingen.de, Presently at ITP, Univ. Goettingen}
\affiliation{Indian Institute of Science Education and
Research-Kolkata, Mohanpur 741252, India}
\author{Sushanta Dattagupta}
\email{sushantad@gmail.com}
\affiliation{ Indian Institute of Science Education and
Research-Kolkata, Mohanpur 741252, India}
\date{\today}
\begin{abstract}
The qubit (or a system of two quantum dots) has become a standard paradigm for studying quantum information processes. Our focus is Decoherence due to interaction of the qubit with its environment, leading to noise. We consider quantum noise generated by a dissipative quantum bath. A detailed comparative study with the results for a classical noise source such as generated by a telegraph process, enables us to set limits on the applicability of this process vis a vis its quantum counterpart, as well as lend handle on the parameters that can be tuned for analyzing decoherence. Both Ohmic and non-Ohmic dissipations are treated and appropriate limits are analyzed for facilitating comparison with the telegraph process. 
\end{abstract}
\maketitle

\section{Introduction }
Environment-induced decoherence is a vexed issue in confronting the challenges to quantum information processes. Simple model calculations thus assume significance, to grasp the underlying parameter-regimes that can be manipulated, in order to minimize the effect of the environment. In quantum optics and solid state physics, the device that has gained popularity in recent times is a qubit (which can be approximately represented by a pair of two quantum dots)$^{2,3}$. This system is equivalent to a superconducting Josephson junction$^{4}$ or a spin-1/2 NMR nucleus$^{5}$. A qubit can be described by a two-level Hamiltonian, the quantum mechanics of which is rather straightforward. Environmental influences can also be easily incorporated in this Hamiltonian. There have been two distinctive attempts in modeling the environment, either in terms of ($i$) a classical stochastic process, such as a Gaussian or a telegraph one, or ($ii$) an explicit quantum collection of bosonic oscillators. One of the main objectives in this paper is to seek a unification of these two apparently disparate approaches. 

We consider an electron hopping between two single level quantum dots, coupled via a tunneling coefficient $\Delta$ and an asymmetric bias $\epsilon$. Denoting by $|L\rangle$ and $|R\rangle$ the left and right dot states, the qubit Hamiltonian, indicated by the subscript $q$, can be written as,
\begin{equation}
 \mathcal{H}_q=\Delta \left(|L\rangle \langle R|+|R\rangle \langle L|\right)+\epsilon\left(|L\rangle \langle L|-|R\rangle \langle R|\right)
\end{equation}
If we introduce the so-called bonding and antibonding states as the spin 'up' and spin 'down' states of the Pauli matrix  $\sigma_z$, denoted by $|+\rangle$ and $|-\rangle$ respectively, as

\begin{eqnarray}
|+\rangle=\frac{1}{\sqrt{2}}\left(|L\rangle+|R\rangle\right), \nonumber \\
|-\rangle=\frac{1}{\sqrt{2}}\left(|L\rangle-|R\rangle\right).
\end{eqnarray}
$\mathcal{H}_q$ can be cast into more familiar notation:

\begin{equation}
\mathcal{H}_{q}=\Delta\sigma_{z}+\epsilon\sigma_{x},
\end{equation} 
where $\sigma_x$ is the other Pauli matrix that is off-diagonal in the representation in which $\sigma_z$ is diagonal.

In order to incorporate the effect of the environment on the qubit our strategy is to expand the Hilbert space to re-express the system Hamiltonian as 
\begin{equation}
\mathcal{H}_{s}=\mathcal{H}_{q}+\tau_{z}(\zeta_{\Delta}\sigma_{z}+\zeta_{\epsilon}\sigma_{x}),
\end{equation}
where $\tau_z$ is yet another Pauli matrix, which could represent a spin-1/2 impurity, for instance. The idea is to couple the $\tau_z$-system to a bath of bosonic oscillators, thereby causing fluctuations in $\tau_z$. The resulting full Hamiltonian can be written as
\begin{equation}
\mathcal{H}=\mathcal{H}_s+\tau_{x}\sum_{k}g_{k}(b_{k}+b_{k}^{\dagger})+\sum_{k}\omega_{k}b_{k}^{\dagger}b_{k}.\end{equation}
Here $b_k^{\dagger}$($b_k$) are boson creation (annihilation) operators, $g_k'^s$ are coupling constants and $\omega_k$ is the harmonic oscillator frequency of the $k^{th}$ mode. 

It is interesting to physically assess the effect of the coupling term (proportional to $g_k$) on the system Hamiltonian $\mathcal{H}_s$. Because $\tau_x$ is purely off-diagonal in the $\tau_z$- representation, it would cause 'spin-flips' in $\tau_z$ between two allowed values +1 and -1, as in the celebrated Glauber model of Ising kinetics$^6$. These flips would occur with an amplitude field that would be proportional to $g_k$ and time-varying bosonic operators $b_k^{\dagger}(t)(b_k(t))$, in the interaction representation of the last term in Eq. (5). The net effect on $\mathcal{H}_s$ is a 'quantum noise', encapsulated by $\tau_z(t)$. 
In a suitable limit, when the latter, i.e. $\tau_z(t)$, could be replaced by a classical noise $\eta(t)$ that jumps at random between $\pm 1$, the system Hamiltonian would be stochastic, given by,
\begin{equation}
\mathcal{H}_{s}(t)=\mathcal{H}_{q}+\eta(t)(\zeta_{\Delta}\sigma_{z}+\zeta_{\epsilon}\sigma_{x}),
\end{equation}
where $\eta(t)$ is a two-state jump process or a Telegraph Process. While we will present below a fully quantum mechanical treatment of Eq. (5), our aim will also be to set conditions under which a classical stochastic description via a telegraph process of $\eta(t)$, as expounded in detail by Tokura and Itakura$^7$, would ensue from the quantum formulation.

The form of the Hamiltonian in Eq. (5) falls under the general scheme of a system-plus-bath approach to nonequilibrium statistical mechanics, in which the full Hamiltonian is written as,
\begin{equation}
\mathcal{H}=\mathcal{H}_{s}+\mathcal{H}_{I}+\mathcal{H}_{B},
\end{equation}
where $\mathcal{H}_I$ is the interaction Hamiltonian and $\mathcal{H}_B$ is the bath Hamiltonian. In the present instance, 
\begin{eqnarray}
\mathcal{H}_I&=&\tau_{x}\sum_{k}g_{k}(b_{k}+b_{k}^{\dagger}), \nonumber \\
\mathcal{H}_B&=&\sum_{k}\omega_{k}b_{k}^{\dagger}b_{k}.
\end{eqnarray}
As we will be employing a Liouvillean approach to the dynamics governed by $\mathcal{H}$, it is useful to explain the notation. The Liouville operator $\mathcal{L}$, associated with $\mathcal{H}$, is defined by
\begin{equation}
 \mathcal{L}A=\frac{1}{\hbar}\left[\mathcal{H},A\right],
\end{equation}
where $A$ is an arbitrary but ordinary operator. Thus a Liouville operator yields an ordinary operator when it operates on an operator (as in the right hand side of Eq. (9)), just as an operator, operating on a wavefunction, yields another wavefunction. We will designate the 'states' of a Liouville operator by round brackets: $|)$, in analogy with the Dirac ket vectors: $|\rangle$ for the wavefunctions. The decomposition in Eq. (7) further allows $\mathcal{L}$ to be split as 
\begin{equation}
 \mathcal{L}=\mathcal{L}_s+\mathcal{L}_I+\mathcal{L}_B.
\end{equation}

The Liouville dynamics effected by $\mathcal{L}$ on the density operator $\rho$, leading to a non-Markovian master equation, in the context of the decoherence of a qubit (as well as a collection of qubits) have been extensively studied by Kurizki and collaborators$^9$. They have also considered the additional influence of external time-varying fields for controlling decoherence, dynamically. While we will be employing a very similar approach, our treatment will differ from Gordon etal$^9$. The latter utilize the Born approximation, valid for weak coupling constants $g_k$, whereas we will consider strong coupling in which $g_k$ will be treated to all orders.

Such strong coupling considerations are the hallmark of functional integral treatments of quantum dissipative systems eg. that of a spin-boson Hamiltonian$^{10}$. Furthermore our focus in the classical limit of the quantum noise will be a telegraph process as opposed to the much-studied Gaussian stochastic processes. Similar spin-boson Hamiltonians, akin to Eq. (5), have been considered in the past$^{11}$, though the comparison has not been dealt with, to the best of our knowledge.  

With these preliminaries the paper is section wise organised as follows. In Sec. II we outline the mathematical steps for a resolvent expansion of the averaged time-development operator in the Laplace transform space, the average being carried out over the Hilbert spaces of $\tau_z$ and $\mathcal{H}_B$. The latter is expressed in terms of a 'self-energy' whose form is provided. In Sec. III explicit results are presented for the much explored Ohmic-dissipation model and comparisons are drawn with the classical telegraph process, results of which are given in the Appendix. For facilitating comparison we discuss fluctuation in bias and hopping separately. In a subsection of Sec. III, we indicate results for non-Ohmic situations as well. Finally in Sec. IV, we discuss the issue of partial decoherence, recently studied by us in the context of the telegraph process$^{12}$, to put it under the perspective of quantum noise. The section V contains a few concluding remarks.

\section{Theoretical Formulation}

In order to incorporate strong-coupling effects it is convenient to transform the Hamiltonian $\mathcal{H}$ in Eq. (5) with the aid of a unitary transformation defined by the operator:
\begin{equation}
S=\exp\left[-\sum_{q}\frac{g_{q}}{2\omega_{q}}\left(b_{q}-b_{q}^{\dagger}\right)\tau_{z}\right]\exp\left[-i\frac{\pi}{2}\tau_y\right].
\end{equation}
The transformed Hamiltonian is given by
\begin{equation}
 \tilde{\mathcal{H}}=S\mathcal{H}S^{-1}.
\end{equation}
What the second term in $S$ does is to cause a rotation in the $\tau$-space by an angle of $\pi/2$ about the y-axis, in the anti-clockwise direction such that $\tau_x\rightarrow\tau_z$ and $\tau_z\rightarrow -\tau_x$. The first term in $S$
then eliminates the interaction term $\mathcal{H}_I$ but puts the onus of coupling on $\mathcal{H}_s$ itself (besides generating an innocuous 'counter term' that is constant, and can be dropped). The net result is

\begin{equation}
\tilde{\mathcal{H}}=\mathcal{H}_q-\frac{1}{2}\left(\tau^{-}B_{+}+\tau^{+}B_{-}\right)\left(\zeta_\Delta\sigma_z+\zeta_\epsilon\sigma_x\right)+\sum_{k}\omega_{k}b_{k}^{\dagger}b_{k},
\end{equation}
where
\begin{eqnarray}
 \tau_\pm&=&\tau_x\pm i\tau_y, \nonumber \\
B_{\pm}&=&\exp\left[\pm\sum_{q}\frac{g_{q}}{2\omega_{q}}\left(b_{q}-b_{q}^{\dagger}\right)\right].
\end{eqnarray}
The transformed Hamiltonian $\mathcal{H}$ is now endowed with a new interaction term that may be written as 
\begin{equation}
 \tilde{\mathcal{H}_I}=-\frac{1}{2}\left(\tau^{-}B_{+}+\tau^{+}B_{-}\right)\left(\zeta_\Delta\sigma_z+\zeta_\epsilon\sigma_x\right).
\end{equation}
The important point however is that any perturbation treatment of $\tilde{\mathcal{H}_I}$ is tantamount to treating the coupling to all orders as $g_q$ now occurs in the exponent as is evident from the structure of $B_\pm$ (cf. Eq. (14)). The equation (13) is again of the generic form of Eq. (7) except that the 'system' is now the qubit itself as $\mathcal{H}_s$ is replaced by $\mathcal{H}_q$!

With reference to the Liovillean in Eq. (10) the equation of motion for the density operator $\rho$ reads
\begin{equation}
\frac{\partial}{\partial t}\rho(t)=-i\mathcal{L}\rho(t). 
\end{equation}
Unlike other approaches$^{7,9}$ we find it convenient to work with the Laplace transforms, defined by
\begin{equation}
 \rho(z)=\int_{0}^{\infty}dt e^{-zt}\rho(t).
\end{equation}
From Eq. (16) then 
\begin{eqnarray}
 \rho(z)&=&\left(U(z)\right)\rho(t=0), \nonumber \\
U(z)&=&\left(z+i\mathcal{L}\right)^{-1}.
\end{eqnarray}
We employ the usual factorization approximation that at $t=0$ the qubit and the bath are decoupled$^8$, so that
\begin{equation}
 \rho(t=0)=\rho_q(0)\otimes\rho_B,
\end{equation}
where $\rho_q(0)$ is the qubit density operator at time $t=0$, $\rho_B=\exp(-\beta\mathcal{H}_B)/Z_B$, is the canonical density operator for the bath and $Z_B$ is the corresponding partition function. Underlying the prescription in Eq. (19) is the assumption that the bath always remains in quilibrium at a fixed temperature T while the qubit evolves from an arbitrary state. For most of our derived expressions we shall assume that the electron is on the left dot at $t=0$ i.e.
\begin{equation}
 \rho_q(0)=|L\rangle\langle L|.
\end{equation}
 Translated to the 'bonding' and 'anti-bonding' basis, this implies
\begin{equation}
 \rho_q(0)=\frac{1}{2}\left(1+\sigma_x\right).
\end{equation}
(We shall return in Sec. IV to a more general initial condition.) The Laplace transform of the so-called reduced density operator can be written as 
\begin{equation}
 \rho_\mathcal{R}(z)={Tr_\tau\otimes Tr_B\left[\left(U(z)\right)\rho_B\right]}\rho_q(0),
\end{equation}
where $Tr_\tau(...)$ denotes trace over the eigenstates of $\tau_z$ whereas $Tr_B(...)$ specifies trace over the bath (i.e. eigenstates of $\mathcal{H}_B$). Because the trace is invariant under the unitary transformations (eg. in Eq. (11)), Eq. (22) can be equivalently expressed as
\begin{equation}
 \rho_\mathcal{R}(z)={Tr_\tau\otimes Tr_B\left[\left(\tilde{U}(z)\right)\rho_B\right]}\rho_q(0),
\end{equation}
where the tilde on $U(z)$ implies that the corresponding time-evolution is now governed by $\tilde{U}$ of Eq. (13).

Denoting the Liouville operator associated with $\tilde{\mathcal{H}}_I$ as $\tilde{\mathcal{L}}_I$, we have upto second order in $\tilde{\mathcal{H}}_I^8$,
\begin{equation}
 \rho_R(z)=\left[\left(z+i\mathcal{L}_q+\tilde{\sum}(z)\right)^{-1}\rho_q(0)\right],
\end{equation}
  where the self-energy $\tilde{\sum}(z)$ is 
\begin{equation}
 \tilde{\sum}(z)=Tr_\tau\otimes {Tr_B\left[\tilde{\mathcal{L}}_I\left(z+i(\mathcal{L}_q+\mathcal{L}_B)\right)^{-1}\tilde{\mathcal{L}}_I\rho_B\right]}.
\end{equation}
The z-dependence (or frequency dependence) of $\tilde{\sum}(z)$ implies that the underlying dynamics is non-Markovian, in general. We emphasize once more that though the self-energy is computed to second order in $\tilde{\mathcal{L}}_I$, the theory is valid to arbitrary orders in the original coupling constants $g_k$. Hence, the resultant master equation in the time domain that emanates from Eq. (24) is of more general validity than the one obtained in the Born-approximation$^9$.

Our next step is to write the matrix elements of the self-energy $\tilde{\sum}(z)$. It is clear that $\tilde{\sum}(z)$ is a Liouvillean in the $\sigma$-subspace alone because the $\tau$ and the bath variables are averaged over, in Eq. (25). Hence, $\tilde{\sum}(z)$ has a $4\times 4$ matrix representation in the 'eigen basis' $|\mu\nu)$ (note the round brackets used for 'states' of the Liouvillean), wherein $|\mu\rangle$, $|\nu\rangle$,... are the eigenkets of $\sigma_z$. Clearly then, the rows and columns of the $4\times 4$ matrix would be labelled by $|++)$, $|--)$, $|+-)$ and $|-+)$, respectively. 
Additionally, while performing the traces in Eq. (25), we would need to sum over the eigenstates of $\tau_z$ and $\mathcal{H}_B$, as in the following:
\begin{eqnarray}
 \tau_z|s_1\rangle&=&s_1|s_1\rangle, \nonumber \\
\mathcal{H}_B|n_b\rangle&=&E_{n_b}|n_b\rangle,
\end{eqnarray}
 etc, With these explanatory remarks about the notation, the matrix elements of $\tilde{\sum}(z)$ turn out to be generalized forms of those given in Ref. [13]:

\begin{widetext}
\begin{eqnarray}
(\mu\nu|\tilde{\sum}(z)|\mu'\nu')=\sum_{n_{b}n_{b}^{1}s_{1}s_{2}}&[&\delta_{\nu\nu'}\sum_{\eta}\frac{\left\langle \mu n_{b}s_{1}\right|H_{I}\left|\eta n_{b}^{1}s_{2}\right\rangle \left\langle \eta n_{b}^{1}s_{2}\right|H_{I}\left|\mu'n_{b}s_{1}\right\rangle }{z+i\left(E_{\eta}-E_{\nu}\right)+i\left(E_{n_{b}^{1}}-E_{n_{b}}\right)}\nonumber \\
&+&\delta_{\mu'\mu}\sum_{\eta}\frac{\left\langle \nu'n_{b}s_{1}\right|H_{I}\left|\eta n_{b}^{1}s_{2}\right\rangle \left\langle \eta n_{b}^{1}s_{2}\right|H_{I}\left|\nu n_{b}s_{1}\right\rangle }{z+i\left(E_{\mu}-E_{\eta}\right)+i\left(E_{n_{b}}-E_{n_{b}^{1}}\right)}\nonumber \\
&-&\frac{\left\langle \mu n_{b}s_{1}\right|H_{I}\left|\mu'n_{b}^{1}s_{2}\right\rangle \left\langle \nu'n_{b}^{1}s_{2}\right|H_{I}\left|\nu n_{b}s_{1}\right\rangle }{z+i\left(E_{\mu'}-E_{\nu}\right)+i\left(E_{n_{b}^{1}}-E_{n_{b}}\right)}\nonumber \\
&-&\frac{\left\langle \mu n_{b}s_{1}\right|H_{I}\left|\mu'n_{b}^{1}s_{2}\right\rangle \left\langle \nu'n_{b}^{1}s_{2}\right|H_{I}\left|\nu n_{b}s_{1}\right\rangle }{z+i\left(E_{\mu}-E_{\nu'}\right)+i\left(E_{n_{b}}-E_{n_{b}^{1}}\right)}]\langle n_b|\rho_B|n_b\rangle.
\end{eqnarray}

From the structure of $\tilde{\mathcal{H}}_I$ in Eq. (15) it is evident that the bath variables enter only through the operators $B_\pm$. Therefore, when we sum over the bath indices $n_b$, $n_b'$, etc., with the Boltzmann weight $\langle n_b|\rho_B|n_b\rangle$, we end up with correlation fucntions of $B_\pm$, as in the following:

\begin{equation}
\Phi_{ij}(t)=Tr_B\left(\rho_B B_i(0)B_j(t)\right) = \langle B_i(0)B_j(t)\rangle \quad(i,j=+ or -).
\end{equation}
As it happens, the only non-zero components of the correlation function are

\begin{eqnarray}
 &&\Phi_{+-}(t)=\Phi_{-+}(t)= \langle B_\pm(0)B_\mp(t)\rangle \nonumber \\
&&=\exp\left\{-\sum_{q}\frac{4g_{q}^2}{\omega_{q}^2}\left[\coth{\left(\frac{\beta\omega_q}{2}\right)}\left(1-\cos\omega_qt\right)+i\sin\omega_qt\right]\right\}.
\end{eqnarray}

An essential attribute of quantum dissipative system is that the system $\mathcal{H}_B$ is taken to the limit of an infinitely large number of bosonic modes. This implies that the discrete sum over $q$ has to be replaced by an integral over continously varying frequencies $\omega$ with an appropriate weight called the 'spectral density' $J(\omega)$, thus

\begin{equation}
 \phi(\tau)=\exp\left\{-2\int_{0}^{\infty}d\omega \frac{J(\omega)}{\omega^2}\left[\coth{\left(\frac{\beta\omega}{2}\right)}\left(1-\cos\omega\tau\right)+i\sin\omega\tau\right]\right\},
\end{equation}
where 
\begin{equation}
 J(\omega)=2\sum_{q}g_{q}^2\delta(\omega-\omega_q).
\end{equation}
\end{widetext}

\section{Explicit Results and Comparison with Telegraph Noise}

\subsection{Ohmic Dissipation}

In the literature on dissipative quantum systems one model for the spectral density $J(\omega)$ that has received the maximum amount of attention is what gives rise to Ohmic dissipation$^{14}$. In this, $J(\omega)$ is assumed to have a linear frequency dependence with an exponential cut-off $\omega_c$
\begin{equation}
 J(\omega)=K\omega\exp\left(-\frac{\omega}{\omega_c}\right),
\end{equation}
 where $K$ is a phenomenological damping parameter that measures the strength of the coupling with the heat bath. This form allows for an analytic expression for the Laplace transform as defined in Eq. (17), for $\beta\omega_c>>1$ as 
\begin{equation}
\tilde{\phi}(z)=\frac{e^{i\pi K}}{\omega_c}\left(\frac{2\pi}{\beta\omega_c}\right)^{(2K-1)}\frac{\Gamma{(1-2K)}\Gamma{(K+z\beta/2\pi)}}{\Gamma{(1-K+z\beta/2\pi)}}.
\end{equation}
where $\Gamma(..)$ denotes gamma functions.

The significance of the condition $\beta\omega_c>>1$ can be ascertained from the fact that the quantity $\hbar\tau_q=\beta$ (Note: $\hbar$ has been set to unity, hitherto) sets the time scale over which quantum coherence is maintained. Thus, for all frequencies greater than the cut-off $\omega_c$, quantum fluctuations remain strictly coherent. Before we take up the comparison with the telegraph noise it is pertinent to point out that it is meaningful to think of a quantum bath as classical only in the incoherent regime i.e. over time scales larger than $\tau_q$. In other words, the classical limit of the bath obtains for frequencies $\omega<<1/\tau_q$, i.e. the temperature is significantly large, but not so large as to violate the condition $\omega_c\tau_q>>1$, if we want to restrict our discussion within the domain of validity of the analytical result of Eq. (33). If the restriction $\omega\tau_q<<1$ does apply, the frequency or z-dependence in the arguments of the Gamma functions in Eq. (33) can be dropped yielding
\begin{equation}
\lambda=\tilde{\phi}(z=0)=\frac{e^{i\pi K}}{\omega_c}\left(\frac{2\pi}{\beta\omega_c}\right)^{(2K-1)}\frac{\Gamma{(1-2K)}\Gamma{(K)}}{\Gamma{(1-K)}}.
\end{equation}
The parameter $\lambda$ will turn out later to be related to the jump rate of the telegraph process.

With this background we are ready to provide explicit results for the different elements of the density matrix in order to examine decoherence as well as to draw comparison with the results for the telegraph noise. Because the latter have been extensively studied recently by Itakura and Tokura$^7$, we will follow their lead in ignoring the bias term in the original qubit Hamiltonian $\mathcal{H}_q$, i.e. set $\epsilon=0$ though it should be evidently clear that the formalism can easily embrace asymmetric cases ($\epsilon\neq 0$) as well. With this, the relevant Hamiltonian for further considerations can then be written as (cf., Eqs. (3) and (13))

\begin{equation}
\tilde{\mathcal{H}}=\Delta\sigma_z-\frac{1}{2}\left(\tau^{-}B_{+}+\tau^{+}B_{-}\right)\left(\zeta_\Delta\sigma_z+\zeta_\epsilon\sigma_x\right)+\sum_{k}\omega_{k}b_{k}^{\dagger}b_{k}
\end{equation}
For ease of discussion and for remaining close to the treatment of Itakura and Tokura we take up the cases of fluctuation in bias ($\zeta_\Delta=0$) and fluctuation in hopping ($\zeta_\epsilon=0$) separately below.

\subsection{Fluctuation in Bias ($\zeta_\Delta=0$)}

The appropriate interaction Hamiltonian that has to be substituted in th expression for the self energy $\sum(z)$ in Eq. (25) is now given by

\begin{equation}
 \tilde{H}_I=-\frac{1}{2}\zeta_\epsilon\sigma_x\left(\tau^{-}B_{+}+\tau^{+}B_{-}\right)
\end{equation}

Carrying out the summations implied in Eq. (25) we arrive at
\begin{equation}
\tilde{\sum}(z)=\left[\begin{array}{cccc}
a_1(z) & -a_1(z) & 0 & 0\\
-a_2(z) & a_{2}(z) & 0 & 0\\
0 & 0 & a_3(z) & -a_3(z)\\
0 & 0 & -a_3(z) & a_3(z)\end{array}\right],\end{equation}
where
\begin{eqnarray}
 a_1(z)&=&2\zeta_{\epsilon}^{2}[\tilde{\phi}(z_{+})+\tilde{\phi}'(z_{-})], a_2(z)=2\zeta_{\epsilon}^{2}[\tilde{\phi}(z_{-})+\tilde{\phi}'(z_{+})]  \nonumber \\
a_3(z)&=&2\zeta_{\epsilon}^{2}[\tilde{\phi}(z)+\tilde{\phi}'(z)], \quad z_\pm=z\pm2i\Delta.
\end{eqnarray}
Here $\tilde{\phi}(z)$ and $\tilde{\phi}'(z)$ are the Laplace transforms of the correlation functions $\phi(t)$ and $\phi(-t)$, respectively.

The matrix element of $\left(z+i\mathcal{L}_q+\tilde{\sum}(z)\right)$, required in Eq. (24), is obtained from a combination of Eq. (27) and the definition of the matrix elements of the Liouville operator $\mathcal{L}_q$ a la Eq. (A.7). The resultant matrix is of a block-diagonal form and can be easily inverted to yield
\begin{eqnarray}
 \left(\rho_\mathcal{R}(z)\right)_{++}&=&\left(\rho_\mathcal{R}(z)\right)_{--}=\frac{1}{2z}, \nonumber \\
\left(\rho_\mathcal{R}(z)\right)_{+-}&=&\frac{z+2i\Delta+4\zeta_\epsilon^2\left(\tilde{\phi}(z)+\tilde{\phi}'(z)\right)}{2\left[z\left(z+4\zeta_\epsilon^2\left(\tilde{\phi}(z)+\tilde{\phi}'(z)\right)\right)+4\Delta^2\right]}.
\end{eqnarray}
\begin{figure}[h!]
 \centering
  \includegraphics[scale=0.50]{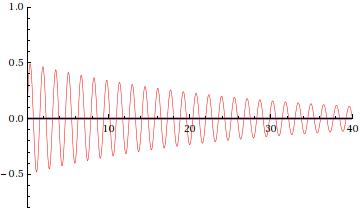}  
 \caption{ $Im[\left(\rho_\mathcal{R}\right)_{LR}(t)]$ is plotted for $\beta=100$,  $k=0.45$ and $\Delta=2$}
\end{figure}
\begin{figure}[h!]
 \centering
  \includegraphics[scale=0.50]{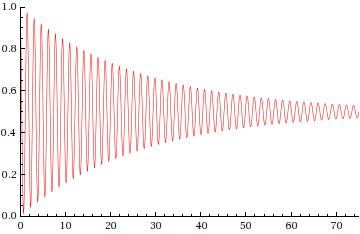} 
 \caption{ $\left(\rho_\mathcal{R}\right)_{LL}(t)$ is plotted for $\beta=100$,  $k=0.45$ and $\Delta=2$}
\end{figure}
\begin{figure}[h!]
 \centering
  \includegraphics[scale=0.50]{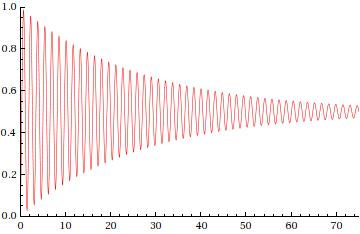} 
 \caption{ $\left(\rho_\mathcal{R}\right)_{RR}(t)$ is plotted for $\beta=100$,  $k=0.45$ and $\Delta=2$}
\end{figure}
 
It is of course possible to go back to the dot-basis with the aid of Eq. (2). Results are shown in Figs. 1, 2 and 3 in which $Im[\rho_{LR}(t)]$, $\rho_{LL}(t)$ and $\rho_{RR}(t)$ are plotted versus time t for a certain choice of parameters. While $Im[\rho_{LR}(t)]$ oscillates and decays to zero as $t\rightarrow\infty$, the diagonal elements, proportional to the populations of the two dot states, oscillate and settle to the common value of 0.5, as we expect for full decoherence. If we compare Figs 1-3 with Fig A.1 for the telegraph process below, it is evident that the quantum case exhibits persistence of oscillations over longer time scales. 

\subsection{Fluctuation in Hopping ($\zeta_\epsilon=0$)}

The appropriate interaction Hamiltonian that has to be substituted in the expression for the self energy $\tilde{\sum}(z)$ in Eq. (25) is now given by

\begin{equation}
 \tilde{H}_I=-\frac{1}{2}\zeta_\epsilon\sigma_z\left(\tau^{-}B_{+}+\tau^{+}B_{-}\right)
\end{equation}

Following the formalism developed in Sec. II and using Eq. (25), the self energy matrix $\tilde{\sum}(z)$ is given by
\begin{equation}
\tilde{\sum}(z)=\left[\begin{array}{cccc}
0 & 0 & 0 & 0\\
0 & 0 & 0 & 0\\
0 & 0 & b_3(z) & 0\\
0 & 0 & 0 & b_4(z)\end{array}\right],\end{equation}
where
\begin{eqnarray}
 b_3(z)&=&4\zeta_{\Delta}^{2}[\phi'(z_{+})+\phi(z_{+})],  \nonumber \\
b_4(z)&=&4\zeta_{\Delta}^{2}[\phi'(z_-)+\phi(z_-)], \quad z_\pm=z\pm2i\Delta, 
\end{eqnarray}
where $\phi(z)$ and $\phi'(z)$ have been defined in the earlier section.

Using the form $\tilde{U}(z)=\left(z+i\mathcal{L}_s+\tilde{\sum}(z)\right)^{-1}$, it is evident from the diagonal form of the matrix that the population of the bonding and anti bonding state remains constant for all $t>0$. Using $\tilde{U}(z)$ the matrix elements are given by,
\begin{eqnarray}
\left(\rho_\mathcal{R}\right)_{++}(z)&=&\left(\rho_R\right)_{--}(z)=\frac{1}{2z}, \nonumber \\
\left(\rho_\mathcal{R}\right)_{+-}(z)&=&\frac{1}{2}\left[\frac{1}{z+2i\Delta+b_{3}}\right].
\end{eqnarray}
 
However in the dot basis we expect the usual saturation to the common value of 0.5 for both left and right dot populations. Fig. (4) shows the variation of left and right dot populations with time and also the loss of coherence over time from the decay of $Im(\rho_{LR}(t))$.
\begin{figure}
 %\centering
\vspace{0.5cm}  
\begin{center}
\includegraphics[scale=0.50]{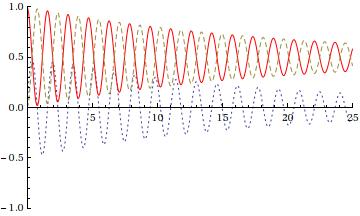} 
\end{center}
 \caption{ $\left(\rho_\mathcal{R}\right)_{LL}(t)$ (solid), $\left(\rho_\mathcal{R}\right)_{RR}(t)$ (dashed) and $Im\left(\rho_\mathcal{R}\right)_{RR}(t)$ (dotted) are plotted for $\beta=100$,  $k=0.45$ and $\Delta=2$}
\end{figure}

We are now in position to make a connection between the results of quantum and classical regimes considered separately in Sec.III and Appendix A. There are two critical distinctions between the two regimes. At the outset, the quantum behavior is non-Markovian, reflected in the z (or frequency)-dependence of the self-energy $\tilde{\sum}(z)$ in Eq. (25). On the other hand, in the classical stochastic case, the Markovian assumption is invoked, to begin with, as seen in Eq. (A.2) below. The second crucial ingredient in the quantum situation, even after the Markovian approximation is incorporated, is in the temperature-dependence of the relaxation rate $\lambda$. While the latter, in the classical domain, is usually endowed with an exponential dependence on temperature, with a barrier-activation in mind, it has a richer structure in the quantum regime, as discussed below.

We mention below in Appendix. A.2 that in the long time region the off-diagonal element of the system density operator follows an  exponential decay with time constant $T_{2}=\frac{1}{\sqrt{ |\lambda^{2}-16\zeta_{\Delta}^{2}|}}$. From Eq. (43), as it can be clearly seen from the Laplace transform of $\rho_{+-}$, the decay is exponential only if we assume $b_{3}(z)$ to be independent of $z$, wherein the time constant is just the inverse of $Re[b_{3}]$. Thus the comparison with the classical telegraph process is meaningful only if one takes the Markovian limit of heat bath induced relaxation, at the outset.

At sufficiently high temperatures given by $k_{B}T>>\zeta_{\Delta}/K$ we can see from Eq. (33) that $\tilde{\phi}(z)$ can be approximated as $\tilde{\phi}(z)=\frac{e^{i\pi K}}{\omega_{c}}\left(\frac{2\pi}{\beta\omega_{c}}\right)^{2K-1}\frac{\Gamma{(1-2K)}\Gamma{(K)}}{\Gamma{(1-K)}}$ which leads to a rather simple expression for $b_{3}$ (for $z=0$) in Eq. (42) as  $b_{3}=\frac{8\zeta_{\Delta}^2}{\omega_{c}}\cos{(\pi K)}(\frac{2\pi}{\beta\omega_{c}})^{2K-1}\frac{\Gamma{(1-2K)}\Gamma{(K)}}{\Gamma{(1-K)}}$, that is real. Hence $\lambda$, which appears as the jump rate in the case of telegraph noise, discussed in the Appendix in the sequel, can be compared with the relaxation rate $Re[b_{3}]$ in the case of quantum noise, is now a function of $K$ and $\beta$. Also, as it is obvious from the expression of the relaxation rate that the dependence on temperature is not an exponential, rather it shows a power law behavior. However, it is pertinent to emphasize that these conclusions hinge on our assumption that quantum dissipation is Ohmic. In the more general case, discussed below in Sec. III D, we will see that the temperature-dependence of the relaxation rate can in fact be exponential, under certain special situations.

At low temperatures and weak damping i.e. 'small' $K$, $\tilde{\phi}(z)$ is given by $\tilde{\phi}(z)=\frac{\mu^2}{z+2\pi/\beta}$ where $\mu$ is a renormalized form of tunneling frequency given by $\mu^2=4\zeta_{\Delta}^2(\frac{2\pi}{\beta\omega_{c}})^{2K-1}\Gamma{(1-2K)}$. Using this form of  $\tilde{\phi}(z)$, and Eq. (43) we can deduce,
\begin{equation}
\left(\rho_\mathcal{R}\right)_{+-}(z)=\frac{1}{2}\left[\frac{1}{z+2i\Delta-\frac{4i\Delta\cos{(\pi K)}\mu^2}{(2\pi/\beta)^2+4\Delta^2}+\frac{4\mu^2(\pi/\beta)\cos{(\pi K)}}{(2\pi/\beta)^2+4\Delta^2}}\right].
\end{equation}
From Eq. (44), it is evident that the relaxation rate becomes $\frac{4\mu^2(\pi/\beta)\cos{(\pi K)}}{(2\pi/\beta)^2+4\Delta^2}$, which is far from an exponential in temperature. Thus it may be stressed that even in the Markovian limit the temperature dependence of both the relaxation rate as well as the Rabi frequency is much more complex than an exponential, as is usually assumed for classical activated processes. 

These conclusions are similar to those obtained earlier in the context of spin relaxation of a muon, tunneling in a double well$^{15}$.

\subsection{Non-Ohmic case}
We had in the text presented results based on the assumption of Ohmic dissipation i.e. choosing $J(\omega)=K\omega\exp{(-\frac{\omega}{\omega_{c}})}$. The Ohmic model is a rather limited one, not applicable to phonons but is relevant to a bath consisting of electron-hole excitations near the Fermi surface, as in metals.$^{14,15}$ In this subsection we will allow for a general spectral density function, and will be analysing the relaxation rate in the case of quantum noise. We will try to make a connection with the classical Telegraph process and show that in the high-temperature limit, the relaxation rate indeed follows an exponential dependence on temperature with however quantum corrections, which was not the case in the Ohmic limit. 
Our starting point is Eq.(30) which upon clubbing the oscillating terms together, can be rewritten as,
\begin{widetext}
\begin{equation}
 \phi(\tau)=\exp\left[-2\int_{0}^{\infty}d\omega \frac{J(\omega)}{\omega^2}\left[\coth{\left(\frac{\beta\omega}{2}\right)}-\frac{\cos{(\omega(\tau-i\beta/2))}}{\sinh{(\frac{\beta\omega}{2})}}\right]\right].
\end{equation}
By making a change of variable from $\tau-i\beta/2$ to $t$ in Eq. (45), $\phi(z)$ can be written as,
\begin{equation}
 \phi(z)= \exp{\left(-\frac{iz\beta}{2}\right)}\int_{-\frac{i\beta}{2}}^{\infty}dt e^{-zt}\exp\left[-2\int_{0}^{\infty}d\omega \frac{J(\omega)}{\omega^2}\left[\coth{\left(\frac{\beta\omega}{2}\right)}-\frac{\cos{(\omega t)}}{\sinh{(\frac{\beta\omega}{2})}}\right]\right],
\end{equation}
In the high temperature regime$^{19}$, the integrand in the second term in Eq. (46), can be replaced by its short time limit as,
\begin{equation}
\phi(z)= \exp{\left(-\frac{iz\beta}{2}\right)}\int_{-\frac{i\beta}{2}}^{\infty}dt e^{-zt}\exp\left[-2\int_{0}^{\infty}d\omega \frac{J(\omega)}{\omega^2}\left[\tanh{\left(\frac{\beta\omega}{4}\right)}+\frac{\omega^2t^2}{2\sinh{(\frac{\beta\omega}{2})}}\right]\right]. 
\end{equation}
Also in the high temperature regime i.e. low $\beta$, $\tanh(x)$ and $\sinh(x)$ can be replaced by the argument i.e. $x$, and hence the integrations in Eq. (47) can be carried out easily to yield,
\begin{equation}
 \phi(z)=\exp{\left(-\beta(a+iz/2)\right)}\int_{-\frac{i\beta}{2}}^{\infty}dt e^{-zt}e^{\frac{4at^2}{\beta}},
\end{equation}
where $a=\frac{1}{2}\int_{0}^{\infty}d\omega \frac{J(\omega)}{\omega}$. Also $\phi'(z)$ is obtained by replacing $i$ by $-i$ in Eq. (48). Thus the relaxation rate which is given by Re($b_{3}$), for the quantum case in Sec. III.C, is obtained by imposing the Markovian limit i.e. putting $z=0$, leading to
\begin{equation}
 b_{3}=4\zeta_{\Delta}^2\left(\exp{\left(-\beta(a-\Delta-\frac{\Delta^2}{4a})\right)}\int_{\frac{-i\beta}{2}+\frac{i\beta\Delta}{4a}}^{\infty}du \exp{\left(-4au^2/\beta \right)}+\exp{\left(-\beta(a+\Delta-\frac{\Delta^2}{4a})\right)}\int_{\frac{i\beta}{2}+\frac{i\beta\Delta}{4a}}^{\infty}du \exp{\left(-4au^2/\beta \right)}\right).
\end{equation}
\end{widetext}
Finally the lower limit in Eq. (49) can be set to zero and thus the integral is just a function of temperature which follows a power law. Thus the dominant behavior of the relaxation rate is an exponential dependence on temperature, which is what we expect at high temperatures. This analysis therefore provides a regime where we can more meaningfully compare the relaxation rate $\lambda$ for the case of classical telegraph process with the Laplace transform of the bath correlation function in the quantum case. Note that the present treatment does not depend on any specific assumption for the spectral density $J(\omega)$ but is more generally couched, irrespective of whether the phonons are acoustic or optic.

\section{The Issue of Partial Decoherence}
In the context of fluctuation in hopping, it is evident that the system Hamiltonian commutes with the interaction Hamiltonian, which means there is no energy exchange between the system and the bath. Such cases are important for studying partial decoherence$^8$. Thus we generalize our discussion to a more general initial state of a qubit (than of Eq. (20)) given by,

\begin{equation}
|\Psi\rangle=\cos{\left(\frac{\theta}{2}\right)}|L\rangle+e^{i\gamma}\sin{\left(\frac{\theta}{2}\right)}|R\rangle.
\end{equation}

In such a scenario it can be established that the density matrix does not 
evolve to a completely mixed state, rather it approches a limiting 
value, which does contain the off-diagonal components. In this 
situtation coherence is not completely lost and the off-diagonal terms give information about the initial state of the system. We note that the initial qubit information can be retrieved, either from 
experiments related to persistent current or by measuring the population 
of states. The former path is not applicable in our case as we have not 
allowed for the existence of an Aharonov-Bohm flux$^8$.

The density matrix (in $|L\rangle$ and $|R\rangle$ basis) at time $t=0$ is 
given from Eq. (50) by,
\begin{equation}
\tilde{\rho_{\mathcal{R}}}(0)=\left[\begin{array}{cc}\cos^{2}(\theta/2) & 
\cos(\theta/2)\sin(\theta/2)e^{-i\gamma}\\\cos(\theta/2)\sin(\theta/2)e^{i\gamma} 
& \sin^{2}(\theta/2)\end{array}\right],
\end{equation}
which can be translated to the bonding-antibonding basis, via the transformation 
given by Eq. (2). The off-diagonal term $\left(\rho_{\mathcal{R}}\right)_{LR}(t\rightarrow \infty)$, which measures coherence, does not go to zero unlike in the cases discussed in Sec.(III.C and A.2). These particular symmetric cases of $\epsilon=0$ and $\zeta_{\epsilon}=0$ show partial coherence which is important for decoherence-free quantum computation protocols.

\begin{figure}[h!]
 \centering
  \includegraphics[scale=0.50]{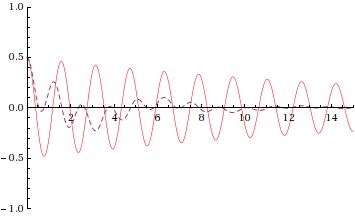} 
 \caption{ $Re[\left(\rho_{\mathcal{R}}\right)_{+-}(t)]$ for quantum (solid) and classical (Dashed) cases are plotted versus time t.}
\end{figure}

\begin{figure}[h!]
 \centering
  \includegraphics[scale=0.50]{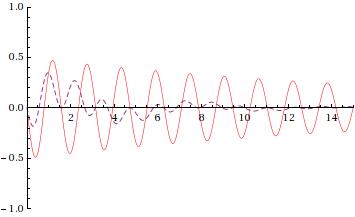} 
 \caption{ $Im[\left(\rho_{\mathcal{R}}\right)_{+-}(t)]$ for quantum (solid) and classical (Dashed) cases are plotted versus time t.}
\end{figure}
 
The above formalism is general and is valid for any kind of environment, as long as the commutator of the system Hamiltonian and the interaction Hamiltonian is zero. As shown in the cases of classical and quantum noises (Figs. 5 and 6), the imaginary and real parts of $\left(\rho_{\mathcal{R}}\right)_{+-}(t)\rightarrow 0$ as $t \rightarrow \infty$, whereas the populations of bonding and anti-bonding states stay constant, as is evident from Eq. (43). 
Thus, in general, the off-diagonal elements of the density matrix (in $|L\rangle$ and $|R\rangle$ basis) approach limiting values, emphasizing that the asymptotic state is not a fully mixed state:

\begin{equation}
\hat{\rho_{\mathcal{R}}}(t\rightarrow \infty) = \left[\begin{array}{cc}\frac{1}{2} & 
Re(\left(\rho_\mathcal{R}\right)_{LR}(0))\\Re(\left(\rho_\mathcal{R}\right)_{RL}(0)) 
& \frac{1}{2}\end{array}\right]. 
\end{equation}

At this stage it is also worthwhile to underscore the importance of time scales involved in attaining partial decoherence. For this we define a function $C(t)$ as:
\begin{equation}
C(t)=\frac{Im(\left(\rho_\mathcal{R}\right)_{LR}(\infty))-Im(\left(\rho_\mathcal{R}\right)_{LR}(t))}{Im(\left(\rho_\mathcal{R}\right)_{LR}(\infty))-Im(\left(\rho_\mathcal{R}\right)_{LR}(0))}.
\end{equation} 
This function is 1 at time $t=0$ and 0 at time $t=\infty$. When integrated over all times, we get a 'relaxation' time for substenance of decoherence as:

\begin{equation}
\tau=\int_{0}^{\infty} dt C(t).
\end{equation}
We carry out an explicit calculation for $\tau$ in both classical and quantum regimes. Using Eq. (A.14) and (43) and the initial density matrix given by Eq. (45), we can calculate $\left(\rho_\mathcal{R}\right)_{LR}(t)$ to obtain:

\begin{equation}
\tau=\frac{\frac{1}{2}I_1+I_2 Im\left(\left(\rho_\mathcal{R}\right)_{LR}(0)\right)}{Im\left(\left(\rho_\mathcal{R}\right)_{LR}(0)\right)},
\end{equation}
where, for the quantum case,
\begin{eqnarray}
I_1=\frac{2\Delta}{(2\Delta)^2+b_{3}^2}, \nonumber \\
I_2=\frac{b_{3}}{(2\Delta)^2+b_{3}^2}, 
\end{eqnarray}
On the other hand, for the classical telegraph process,
\begin{eqnarray}
I_1&=&\frac{\left(2\Delta\right)\left(-\lambda+k\right)}{(2\Delta)^2+\left(\frac{\lambda+k}{2}\right)^2}+\frac{\left(2\Delta\right)\left(\lambda+k\right)}{(2\Delta)^2+\left(\frac{\lambda-k}{2}\right)^2}, \nonumber \\
I_2&=&\frac{\left(\frac{\lambda+k}{2}\right)\left(-\lambda+k\right)}{(2\Delta)^2+\left(\frac{\lambda+k}{2}\right)^2}+\frac{\left(\frac{\lambda-k}{2}\right)\left(-\lambda+k\right)}{(2\Delta)^2+\left(\frac{\lambda-k}{2}\right)^2}, \nonumber \\
k&=&\sqrt{\left(\lambda^2-16\zeta_{\Delta}^2\right)}. 
\end{eqnarray}
where $b_{3}$ is given by the $z=0$ limit of Eq. (42). It is evident from Fig. (6) that the partial decoherence is attained more rapidly in the case of classical noise.
\section{Concluding Remarks}
Our main emphasis in this paper has been a relative assessment of quantum and classical noise sources attached to a pair of quantum dots or a qubit. The analysis has been made with the aid of a unified formalism, comprising a resolvant expansion of an averaged time-development operator. The quantum formalism enabled us to provide a microscopic meaning and detailed temperature-dependence to the phenomenologically introduced parameter of the relaxation rate, that appears in the classical case of a telegraphic noise. Our treatment of the quantum noise has included the much studied ohmic model of dissipation that characterizes electron-hole excitations off the Fermi surface, in a metal, as well as non-ohmic dissipation which covers both acoustic and optic phonons$^{14}$. It is eventually in the phonon model of dissipation that the usually assumed exponential temperature dependence of the relaxation rate (of a Telegraph process) is realized, making the comparison between the classical and quantum cases more direct.

In the last subsection of the paper (see IV) we focussed on the important issue of partial decoherence that can be utilized for quantum computation, which has received recent attention$^8$. This case was analyzed when fluctuation is ascribed only to the hopping term, yielding a situation in which the system Hamiltonian commutes with the coupling to the heat bath. Here, the comparison between classical and quantum noises is quite striking--coherence persists over longer time scales for the quantum case. This attribute can be effectively exploited in the context of quantum computation, in which it is essential to be able to retrieve information about the initial quantum state, notwithstanding heat bath-induced effects. 
\section{Acknowledgement}
EK thanks the INSPIRE support of the Department of Science and Technology for an MS thesis project that contains the present contribution. SD will like to record his gratitude to Amnon Aharony, Ora Entin-Wohlman and Shmuel Gurvitz for many helpful discussions.

\appendix

\addtocontents{toc}{\protect\contentsline{chapter}{Appendix:}{}}
\section{Results for Telegraph Process}
\setcounter{figure}{0}
\subsection{Fluctutaion in Bias ($\zeta_\Delta=0$) under Telegraph Process}

The stochastic Hamiltonian appropriate to Eqs. (35) and (36) is given by (cf., Eq.(6))
\begin{equation}
 \mathcal{H}_s=\Delta\sigma_z+\eta(t)\zeta_\epsilon\sigma_x,
\end{equation}
where $\eta(t)$ is a telegraph process. Concomitantly Eq. (24) is replaced by$^{16}$
\begin{widetext}
\begin{equation}
 \rho_\mathcal{R}(z)=\left[\sum_{a,b}(a|\left(z+i\mathcal{L}_{q}+i\sum_{j=\pm}\mathcal{L}_{j}F_{j}-W\right)^{-1}|b)p_b\right]\rho_q(0),
\end{equation}
\end{widetext}
where the 'stochastic states' $|a)$, $|b)$... are associated with the two possible values $\pm 1$ of $\eta(t)$ (over which the summations in Eq. (A.2) are performed), $p_b$ is the a-priori probability of the occurence of the state $|b)$, $\mathcal{L}_q$ is the Liouville operator accompanying $\mathcal{H}_q=\Delta\sigma_z$, and $\mathcal{L}_j$ is the Liouville operator associated with an ordinary operator $V_j$ (j=+1 or -1) such that 

\begin{equation}
 V_{\pm}=\pm\zeta_\epsilon\sigma_x.
\end{equation}
It remains then to define the 'stochastic matrices' $F_j$ and $W$. The matrix $F_j$ is a projection governed by
\begin{equation}
 \left(a|F_j|b\right)=\delta_{ab}\delta_{aj},
\end{equation}
whereas W is a jump matrix whose element $ (a|W|b)$, for instance, is the rate of jump of the process $\eta(t)$ from the state $|b)$ to the state $|a)$. The underlying Markovian assumption is manifest in the frequency independence of $W$.

The properties of the telegraph process allow us to calculate the average over the stochastic indices a,b,... in Eq. (A.2) in closed form, leading to 
\begin{equation}
 \rho_\mathcal{R}(z)=\left[\frac{\bar{U}_0(z+\lambda)}{1-\lambda \bar{U}_0(z+\lambda)}\right]\rho_q(0),
\end{equation}
where
\begin{equation}
\left(\tilde{U}_{0}(z+\lambda)\right)=\sum_{j}p_{j}\left(z+\lambda+i\mathcal{L}_{q}+i\mathcal{L}_{j}\right)^{-1},\end{equation}
$\lambda$ being the jump rate.

In calculating the matrix elements of $\tilde{U}_{0}(z+\lambda)$ we ned to use the properties of Liouville operators given by$^{16}$
\begin{equation}
 \left(\mu\nu|\mathcal{L}_j|\mu'\nu'\right)=\langle\mu|V_j|\mu'\rangle\delta_{\nu\nu'}-\langle\nu'|V_j|\nu\rangle\delta_{\mu\mu'}.
\end{equation}
Using (A.7), it is straightforward to obtain
\begin{equation}
\left(\bar{z}+i\mathcal{L}_q+i\mathcal{L}_\pm\right)\equiv\left[\begin{array}{cccc}
\bar{z} & \mp i\zeta_{\epsilon} & \pm i\zeta_{\epsilon} & 0\\
\mp i\zeta_{\epsilon} & \bar{z}+2i\Delta & 0 & \pm i\zeta_{\epsilon}\\
\pm i\zeta_{\epsilon} & 0 & \bar{z}-2i\Delta & \mp i\zeta_{\epsilon}\\
0 & \pm i\zeta_{\epsilon} & \mp i\zeta_{\epsilon} & \bar{z}\end{array}\right],
\end{equation}
where 
\begin{equation}
 \bar{z}=z+\lambda
\end{equation}
The matrix of $\bar{U}_0(\bar{z})$ is therefore of a simplified block-diagonal form from which matrix under the square brackets in Eq. (A.5) can be evaluated easily. Combined with the initial condition in Eq. (21) we find, after some straight forward algebra,
\begin{eqnarray}
\left(\rho_\mathcal{R}\right)_{++}(z)&=&\left[\frac{2(a+b)-\lambda( a^{2}-b^{2})}{D_{1}}\right]\left(\rho_\mathcal{R}\right)_{++}(0) \nonumber \\
\left(\rho_\mathcal{R}\right)_{--}(z)&=&\left[\frac{2(a+b)-\lambda( a^{2}-b^{2})}{D_{1}}\right]\left(\rho_\mathcal{R}\right)_{--}(0).
\end{eqnarray}
and
\begin{equation}
 \left(\rho_\mathcal{R}\right)_{+-}(z)=\left[\frac{2(c+b)-\lambda(dd^{*}-b^{2})}{D_{2}}\right]\left(\rho_\mathcal{R}\right)_{+-}(0)
\end{equation}

where
\begin{eqnarray}
 a&=&\frac{2(\bar{z}^2+2\zeta_\epsilon^2+4\Delta^2)}{\bar{z}[\bar{z}^2+4(\zeta_\epsilon^2+\Delta^2)]}, \nonumber \\
b&=&\frac{4\zeta_\epsilon^2}{\bar{z}[\bar{z}^2+4(\zeta_\epsilon^2+\Delta^2)]}, \nonumber \\
c&=&\frac{2[\zeta_\epsilon^2+\bar{z}(\bar{z}-2i\Delta)]}{\bar{z}[\bar{z}^2+4(\zeta_\epsilon^2+\Delta^2)]}, \nonumber \\
D_1&=&(2-\lambda a)^2-(\lambda b)^2, \nonumber \\
D_2&=&(2-\lambda c)(2-\lambda c^*)-(\lambda b)^2.
\end{eqnarray}
\begin{figure}[h!]
 \centering
    \includegraphics[scale=0.60]{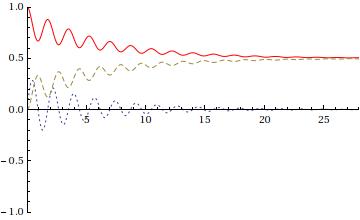}
 \caption{ $\left(\rho_\mathcal{R}\right)_{LL}(t)$ (solid), $\left(\rho_\mathcal{R}\right)_{RR}(t)$ (Dashed), and $Im[\left(\rho_\mathcal{R}\right)_{LR}(t)]$ (Dotted) are plotted versus time t for $\Delta=1$,  $\zeta_{\epsilon}=1.5$ and $\lambda =0.5$}
\end{figure}
With the chosen initial condition in Eq. (21), Eq. (A.10) simplifies to  $\left(\rho_R(z)\right)_{++}=\left(\rho_R(z)\right)_{--}=\frac{1}{2z}.$
It is of course a matter of ease to go from Eqs. (A.10) and (A.11) to the dot basis, with the aid of the transformation in Eq. (2) to rederive the elements of the density operator, results of which are presented in Fig. (A.1). The latter are in complete conformity with those of Ref. [7] and display full decoherence at infinite times i.e. $\left(\rho_\mathcal{R}\right)_{LL}=\left(\rho_\mathcal{R}\right)_{RR}=\frac{1}{2}$ and $\left(\rho_\mathcal{R}\right)_{LR}=0$ as expected.
\subsection{Fluctuation in Hopping ($\zeta_\epsilon=0$) under Telegraph Process}
The relevant stochastic Hamiltonian is now given by (cf., Eq. (6))
\begin{equation}
 \mathcal{H}_s(t)=\Delta\sigma_z+\eta(t)\zeta_\Delta\sigma_z,
\end{equation}
In this case the fluctuating part of the Hamiltonian commutes with the system part of the Hamiltonian, a case which has received attention recently$^{12}$. In this case now there is no energy transfer between the system and the bath, allowing for partial decoherence (see Sec. IV). In this case the calculations are simpler than in A.I as the matrix of $\bar{U}_0(\bar{z})$ is diagonal in the $\sigma_z$ representation. Therefore, the diagonal elememts of $\rho_R(z)$ do not evolve at all from their initial values of one-half whereas the off-diagonal element becomes
\begin{equation}
 \left(\rho_\mathcal{R}(z)\right)_{+-}=\frac{1}{2}\left[z+2i\Delta+\frac{4\zeta_\Delta^2}{\bar{z}+2i\Delta}\right]^{-1}
\end{equation}
The time dependence of the off-diagonal term $\left(\rho_\mathcal{R}(z)\right)_{+-}(t)$ is obtained from the inverse Laplace transform of $\left(\rho_\mathcal{R}(z)\right)_{+-}(z)$ to yield, for $\lambda>4\zeta_{\Delta}$,
\begin{widetext}
\begin{equation}
|\left(\rho_\mathcal{R}(z)\right)_{+-}(t)|=\frac{1}{4\sqrt{\lambda^{2}-16\zeta_{\Delta}^{2}}}\left[\left(\sqrt{\lambda^{2}-16\zeta_{\Delta}^{2}}-\lambda\right)e^{\left(-\frac{1}{2}(\sqrt{\lambda^{2}-16\zeta_{\Delta}^{2}}+\lambda)t\right)}+\left(\sqrt{\lambda^{2}-16\zeta_{\Delta}^{2}}+\lambda\right)e^{\left(-\frac{1}{2}(\lambda-\sqrt{\lambda^{2}-16\zeta_{\Delta}^{2}})t\right)}\right],\label{eq:21-1}\end{equation}
\end{widetext}
This expression coincides with the one derived by Itakura and Tokura$^7$ following an elaborate Dyson series-type time-domain treatment. The offdiagonal term initially follows a Gaussian decay, $|\left(\rho_\mathcal{R}(z)\right)_{+-}(t)|\sim\frac{1}{2}\exp\left\{-2\zeta_{\Delta}^{2}t^{2}\right\}$, discussed in [7], and it becomes exponential in the long time regime. The time constant appearing in the exponential decay is referred to as $T_{2}$ following the NMR literature $^5$, turning out to be $T_{2}=\frac{1}{\sqrt{ |\lambda^{2}-16\zeta_{\Delta}^{2}|}}$ for $\zeta_{\Delta}<<\lambda$. Transforming to the dot basis the behaviour of the elements of the density operator is exhibited in Fig. (A.2).
\begin{figure}[h!]
 \centering
    \includegraphics[scale=0.60]{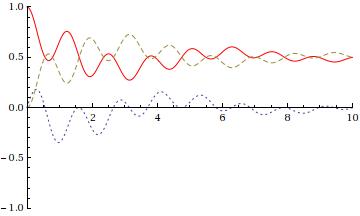}
 \caption{ $\left(\rho_\mathcal{R}(z)\right)_{LL}(t)$ (Solid), $\left(\rho_\mathcal{R}(z)\right)_{RR}(t)$ (Dashed), and $[\left(\rho_\mathcal{R}(z)\right)_{LR}(t)]$ (Dotted) are plotted as a function of time t for $\Delta=1$, $\zeta_{\Delta}=1.5$ and $\lambda =0.5$}
\end{figure}

\end{document}